\documentclass[sigconf]{acmart}

\usepackage{amsthm}
\newtheorem{mydef}{Definition}
\usepackage{multirow}
\usepackage{subcaption}
\usepackage{algorithm}
\usepackage{algorithmic}
\usepackage{enumitem}

\newcommand{\fillblank}[1]{\textcolor{red}{#1}}

\newcommand{\zfcomment}[1]{\textcolor{red}{[#1---zh]}}

\AtBeginDocument{%
  \providecommand\BibTeX{{%
    \normalfont B\kern-0.5em{\scshape i\kern-0.25em b}\kern-0.8em\TeX}}}

\setcopyright{rightsretained}
\acmConference[]{\textnormal{For all other uses}}{\textnormal{and collaborations}}{\textnormal{contact the owner/authors(s).}}
\acmISBN{}
\acmDOI{}

\begin{document}

\title{Learning Post-Hoc Causal Explanations for Recommendation}


\author{Shuyuan Xu$^1$, Yunqi Li$^1$, Shuchang Liu$^1$, Zuohui Fu$^1$, Xu Chen$^2$, Yongfeng Zhang$^1$}
\affiliation{
  \institution{$^1$Department of Computer Science, Rutgers University, New Brunswick, NJ 08901, US}
  \institution{$^2$Gaoling School of Artificial Intelligence, Renmin University of China, Beijing, 100872, China}
}
\email{{shuyuan.xu, yunqi.li, shuchang.liu, zuohui.fu}@rutgers.edu, xu.chen@ruc.edu.cn, yongfeng.zhang@rutgers.edu}


\begin{abstract}
State-of-the-art recommender systems have the ability to generate high-quality recommendations, but usually cannot provide intuitive explanations to humans due to the usage of black-box prediction models. The lack of transparency has highlighted the critical importance of improving the explainability of recommender systems. In this paper, we propose to extract causal rules from the user interaction history as post-hoc explanations for the black-box sequential recommendation mechanisms, whilst maintain the predictive accuracy of the recommendation model. Our approach firstly achieves counterfactual examples with the aid of a perturbation model, and then extracts personalized causal relationships for the recommendation model through a causal rule mining algorithm. Experiments are conducted on several state-of-the-art sequential recommendation models and real-world datasets to verify the performance of our model on generating causal explanations. Meanwhile, We evaluate the discovered causal explanations in terms of quality and fidelity, which show that compared with conventional association rules, causal rules can provide personalized and more effective explanations for the behavior of black-box recommendation models.
\end{abstract}





\maketitle
\pagestyle{plain}

\section{Introduction}

As widely used in decision-making, recommender systems have been recognized for its ability to provide high-quality services that reduces the gap between products and customers.
Nowadays, many state-of-the-art performances are achieved by neural network models.
Typically, deep learning models are used as black-box latent factor models accompanied with a high-dimensional latent space.
This allows them to achieve good expressiveness power and accuracy in various recommendation tasks.
However, it is also true that complex neural models 
easily go beyond the comprehension of the majority of customers, since thousands or millions of parameters are involved.
Nevertheless, it is a natural demand of human-beings to understand why a model makes a specific decision, rather than blindly accepting the results without knowing the underlying reason. As a result, providing supportive information and interpretation along with the recommendation can be helpful for both the customers and the platform,
since it improves the transparency, trustworthiness, effectiveness, and user satisfaction of the recommendation systems, while facilitating system designers to refine the algorithms \cite{zhang2018explainable}. For example, a user may be inspired by the recommendation panel explained as ``you may also like'' in an e-commerce system, and thus decides to look around for more items that better satisfy his or her interests. In the meantime, the user preference would be captured more precisely, so that better services can be provided in further interactions. 
On the other hand, system designers may also easily figure out the reasons of providing unsatisfied recommendation,
leveraging the result to take actions.

\begin{figure}[t]
    \centering
    \includegraphics[scale=0.5]{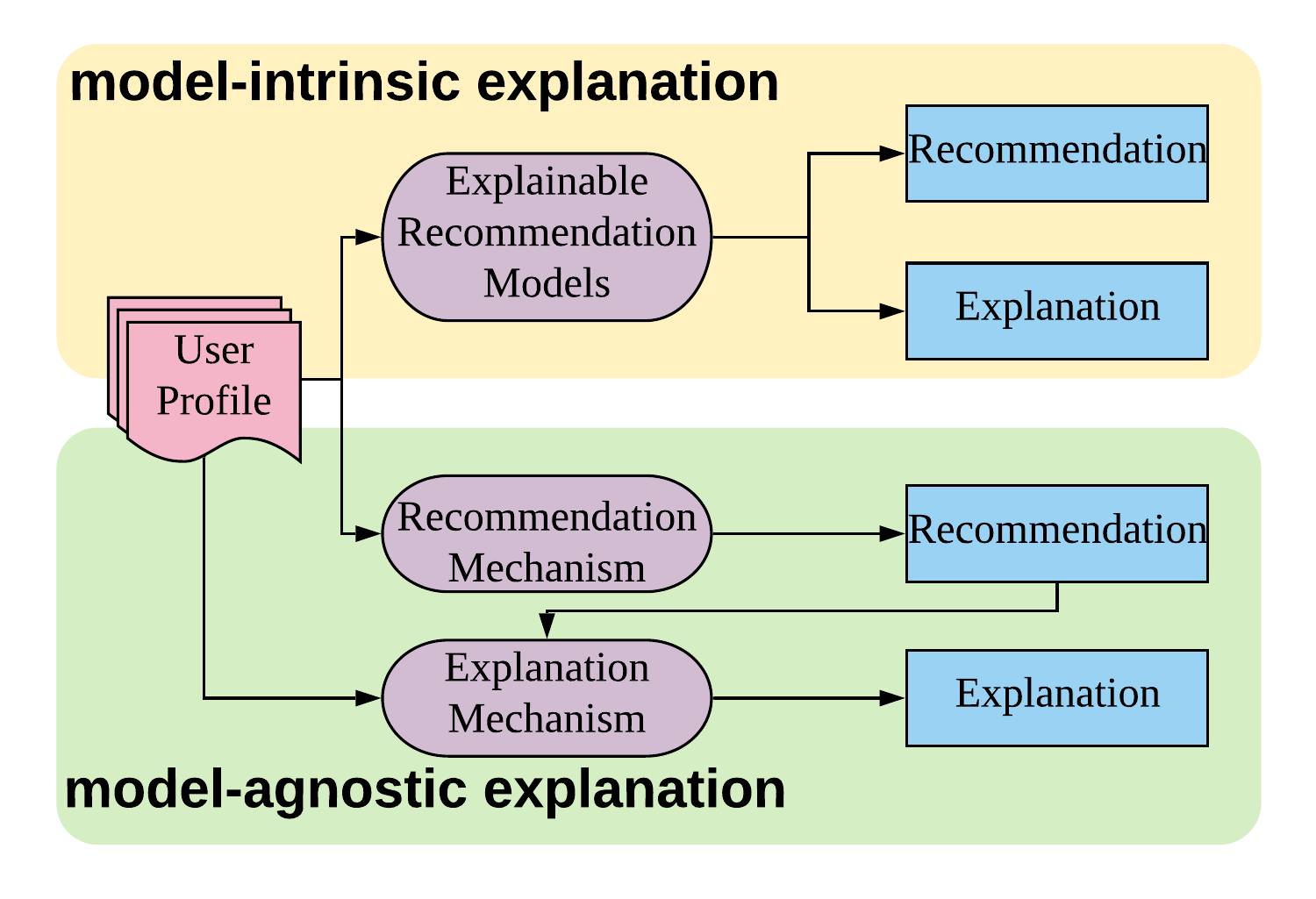}
    \vspace{-15pt}
    \caption{The model-intrinsic approaches directly generate the recommendations and explanations, while the model-agnostic approaches generate explanations after the black-box model has generated the recommendations.}
    \vspace{-20pt}
    \label{fig:posthoc}
\end{figure}

To address the explainability problem, researchers turned to explainable recommendation models, which are expected to not only generate effective recommendations but also intuitive explanations to humans.
Generally, the explainable models can be either model-intrinsic or model-agnostic (also known as post-hoc), as shown in Fig.\ref{fig:posthoc}. 
The model-intrinsic approach takes advantage of the interpretable mechanism of the model, with the explanations directly provided as the intermediate stages of recommended decisions.
For instance, simple user-based collaborative filtering methods 
pass an item from one user to another similar user, and thus it can directly output explanations like ``similar users also bought this item''.
However, the recommendation explanations are usually not obtained for free, and sometimes we have to trade-off with model accuracy~\cite{theodorou2017designing}. In many cases, outstanding recommendation performances are usually achieved by models that are less interpretable.
Thus, it is very challenging for current explainable recommendation methods to redesign a black-box model into an interpretable one whilst maintaining the recommendation performance \cite{zhang2018explainable}.

In contrast, post-hoc models make no assumption of the underlying recommendation model, and allow the decision mechanism to be a black-box, since it will provide explanations after a decision is made. 
Although such explanations may not strictly follow the exact mechanism that generated the recommendations, they offer the flexibility to be applied to many different models.
Though it is still not fully understood what information is useful for generating explanations for a certain recommendation result, Peake ~\cite{peake2018explanation} argued that one can provide post-hoc item-level explanations.
In other words, interacted items (the causes) in a user's history can be used as explanations for the future item recommendations (the effect), since it answers ``what items you bought causes the recommendation of this item''.

However, existing work mostly use global association rule mining to discover the relationship between items, which relies on the item co-occurrence among all user transactions.
Therefore, the explanations are not personalized, i.e., different users would receive the same explanation as long as they are recommended with the same item and have overlapped histories.
This makes it incompatible with modern recommender systems, which aims to provide personalized services to users. 
Moreover, the item-level explanation problem naturally involves causal analysis between a user's previous and future behaviors, which makes the problem even more challenging since it has to answer counterfactual questions such as ``what would happen if a different set of items were purchased''.

In this paper, we explore a causal analysis framework to provide post-hoc causal explanations for any sequential recommendation algorithm.
The goal is to design a model that generates post-hoc explanations for black-box recommendation models in order to reduce the accuracy-interpretability trade-off. Since the explanation model and recommendation model work separately, we obtain the benefit of explainability without hurting the prediction performance. Technically, we propose a Variational Auto-Encoder (VAE) based perturbation framework to create counterfactual examples for causal analysis, which extracts causal rules between a user's previous and future behaviors as explanations.

The key contributions of this paper are as follows:

\begin{itemize}
    \item We design and study a causal rule mining framework for sequential recommendation.
    \item We show that this framework can generate personalized post-hoc explanations based on item-level causal rules to explain the behaviors of a sequential recommendation model.
    \item We conduct experiments on real-world data to show that our explanation model outperforms state-of-the-art baselines.
\end{itemize}

In the following, we review related work in in Section 2, and introduce our model in Section 3. Experimental settings and results are provided in Section 4. We conclude this work in Section 5.

\section{Related Work}
\subsection{Sequential Recommendation}
Sequential recommendation takes into account the historical order of items interacted by a user, and aims to capture useful sequential patterns for making consecutive predictions of the user's future behaviors.
Rendle et al. \cite{rendle2010factorizing} proposed Factorized Personalized Markov Chains (FPMC) to combine Markov chain and matrix factorization for next basket recommendation. 
The Hierarchical Representation Model (HRM) \cite{wang2015learning} further extended this idea by leveraging representation learning as latent factors in a hierarchical model. 
However, these methods can only model the local sequential patterns of very limited number of adjacent records. 
To model multi-step sequential behaviors, He et al. \cite{he2016fusing} adopted Markov chain to provide recommendations with sparse sequences.
Later on, the rapid development of representation learning and neural networks introduced many new techniques that further push the research of sequential recommendation to a new level. 
For example, Hidasi et. al. \cite{hidasi2015session} used an RNN-based model to learn the user history representation,
Yu et. al. \cite{yu2016dynamic} provided a dynamic recurrent model, 
Li et. al. \cite{li2017neural} proposed an attention-based GRU model, Chen et. al. \cite{chen2018sequential} developed user- and item-level memory networks, and Huang et. al. \cite{huang2018improving} further integrated knowledge graphs into memory networks.
However, most of the models exhibit complicated neural network architectures, and it is usually difficult to interpret their prediction results. 
As a result, we would like to generate explanations for these black box sequential recommendation models. 

\subsection{Explainable Recommendation}
Explainable recommendation focuses on developing models that can generate not only high-quality recommendations but also intuitive explanations, which help to improve the transparency of the recommendation systems \cite{zhang2018explainable}. 
Generally, the explainable models can either be model-intrinsic or model-agnostic as introduced in introduction. 
As for model-intrinsic approaches, lots of popular explainable recommendation methods, such as factorization models \cite{zhang2014explicit,chen2018attention,wang2018explainable}, 
deep learning models \cite{seo2017interpretable,chen2018sequential,li2019capsule,costa2018automatic},
and knowledge graph models \cite{ai2018learning,xian2019reinforcement,huang2018improving,fu2020fairness,ma2019jointly,wang2019explainable} 
have been proposed. A more complete review of the related models can be seen in \cite{zhang2018explainable}.
However, they mix the recommendation mechanism with interpretable components, which often results in a system too complex to make successful explanations.
Moreover, the increased model complexity may reduce the interpretability.
A natural way to avoid this dilemma is to rely on model-agnostic post-hoc approaches so that the recommendation system is free from the noises of the down-stream explanation generator. 
Examples include \cite{mcinerney2018explore} that proposed a bandit approach, \cite{wang2018reinforcement} that proposed a reinforcement learning framework to generate sentence explanations, and \cite{peake2018explanation} that developed an association rule mining approach. 
In their work, the transactions of all users, which consider user history as input and recommendation item as output, are used to extract association rules as the explanation for black-box models. 
However, correlation is not reliable for its direction agnostic feature.
As will mention in the next part, our goal here is to find causal relationships in the user behaviour, which we can provide more stable explanations. 

\subsection{Causal Inference in Recommendation}
Originated as statistical problems, causal inference \cite{pearl2000causality,imbens2015causal}
aims at understanding and explaining the causal effect of one variable on another.
While the observational data is considered as the factual world, causal effect inferences should be aware of the counterfactual world, thus often regarded as the question of "what-if".
The challenge is that it is often expensive or even impossible to obtain counterfactual data. 
For example, 
it is immoral to re-do the experiment on a patient to find out what will happen if we have not given the medicine.
Though the majority of causal inference study resides in the direction of statistics and philosophy, it recently attract attention from AI community for its great power of explainablity and bias elimination ability.
Efforts have managed to bring causal inference to several machine learning areas, including recommendation \cite{bonner2018causal}, 
learning to rank \cite{joachims2017unbiased},
natural language processing \cite{wood2018challenges}, 
and reinforcement learning  \cite{buesing2018woulda}, 
etc. 
With respect to recommendation tasks, large amount of work is about how to achieve de-bias matrix factorization with causal inference.
The probabilistic approach ExpoMF proposed in \cite{liang2016modeling} directly incorporated user exposure to items into collaborative filtering, where the exposure is modeled as a latent variable. 
Liang et. al.\cite{liang2016causal} followed to develop a causal inference approach to recommender systems which believed that the exposure and click data came from different models, thus using the click data alone to infer the user preferences would be biased by the exposure data. 
They used causal inference to correct for this bias for improving generalization of recommendation systems to new data. 
Bonner et. al.\cite{bonner2018causal} proposed a new domain adaptation algorithm which was learned from logged data including outcomes from a biased recommendation policy, and predicted recommendation results according to random exposure. 
Differently, this paper focuses on learning causal rules to provide more intuitive explanation for the black-box recommendation models. 
Additionally, we consider \cite{alvarez2017causal} as a highly related work though it is originally proposed for natural language processing tasks.
As we will discuss in the later sections, we utilize some of the key ideas of its model construction, and show why it works in sequential recommendation scenarios.

\vspace{-3pt}
\section{Proposed Approach}
In this section, we first define the explanation problem and then introduce our model as a combination of two parts: a VAE-based perturbation model that generates the counterfactual samples for causal analysis, and a causal rule mining model that can extract causal dependencies between the cause-effect items. 

\subsection{Problem Setting}

We denote the set of users as $\mathcal{U}=\{u_1,u_2,\cdots, u_{|\mathcal{U}|} \}$ and set of items as $\mathcal{I}=\{i_1,i_2.\cdots,i_{|\mathcal{I}|}\}$. 
Each user $u$ is associated with a purchase history represented as a sequence of items $\mathcal{H}^u$.
The $j$-th interacted item in the history is denoted as $H_j^u \in \mathcal{I}$.
A black-box sequential recommendation model $\mathcal{F}:\mathcal{H}\rightarrow\mathcal{I}$ is a function that takes a sequence of items (as will discuss later, it can be the permuted user history) as input and outputs the recommended item.
In practice, the underlying mechanism usually consists of two steps: a ranking function that scores all candidate items based on the user history, and then selects the item with best score as the final output.
Note that it only uses user-item interaction without any content or context information, and the scores predicted by the ranking function may differ according to the tasks (e.g. $\{1,\dots,5\}$ for rating prediction, while $[0,1]$ for CTR prediction).
Our goal is to find an item-level post-hoc model that captures the causal relation between the history items and the recommended item for each user.

\begin{mydef}
(Causal Relation) For two variables $X$ and $Y$, if $X$ triggers $Y$, 
then we say that there is a causal relation $X \Rightarrow Y$, where $X$ is the \textbf{cause} and $Y$ is the \textbf{effect}.
\end{mydef}

When a given recommendation model $\mathcal{F}$ maps a user history $\mathcal{H}^u$ to a recommended item  $Y^u\in\mathcal{I}$, all items in $\mathcal{H}^u$ are considered as potential causes of $Y^u$.
Thus we can formulate the set of causal relation candidates as $\mathcal{S}^u = \{(H,Y^u)|H\in\mathcal{H}^u\}$.

\begin{mydef}
(Causal Explanation for Sequential Recommendation Model) Given a causal relation candidate set $\mathcal{S}^u$ for user $u$, if there exists a true causal relation $(H,Y^u) \in\mathcal{S}^u$, then the causal explanation for recommending $Y^u$ is described as ``Because you purchased $H$, the model recommends you $Y^u$'', denoted as $H\Rightarrow Y^u$.
\end{mydef}



Then the remaining problem is how to determine whether a candidate pair is a true causal relation.
We can mitigate the problem by allowing a likelihood estimation for a candidate pair to be causal relation. 
In other words, we would like to find a ranking function that predicts the likelihood for each candidate pair.
In this way, causal explanations can be generated by selecting the most promising ones from these candidates.
One advantage of this formulation is that it allows the possibility that there is no causal relation between a user's history and the recommended item,
e.g. when algorithm recommends the most popular items regardless of the user history.
In the following sections, we will illustrate in detail how our model solves these problems.

\begin{figure*}[htbp]
    \vspace{-10pt}
    \centering
    \includegraphics[width=\linewidth]{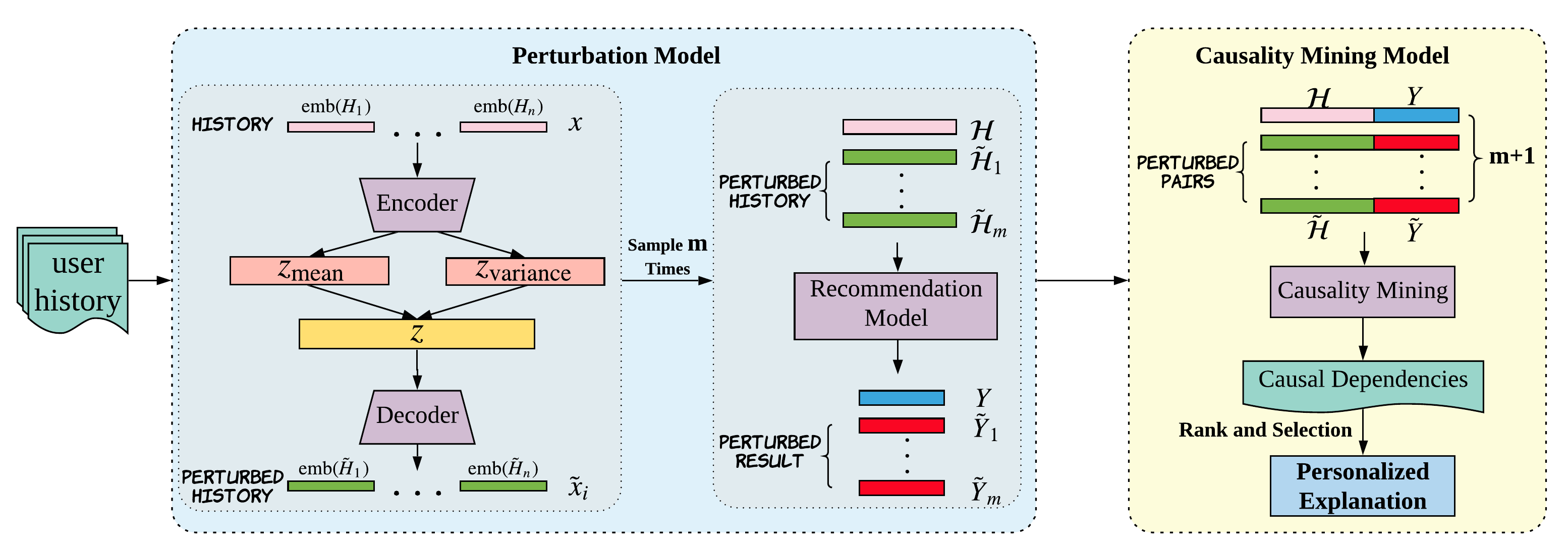}
    \vspace{-20pt}
    \caption{Model framework. $x$ is the concatenation of the item embeddings of the user history. $\tilde{x}$ is the perturbed embedding. 
    }
    \vspace{-10pt}
    \label{fig:model}
\end{figure*}

\begin{algorithm}[t]
    \caption{Causal Post-hoc Explanation Model}
    \begin{flushleft}
    \textbf{Input:} users $\mathcal{U}$, items $\mathcal{I}$, user history $\mathcal{H}^{u}$, perturbation times $m$, black-box model $\mathcal{F}$, embedding model $\mathcal{E}$, causal mining model $\mathcal{M}$\\
    \textbf{Output:} causal explanations $H\Rightarrow Y^{u}$ where $H\in\mathcal{H}^{u}$\\
    \end{flushleft}
    \begin{algorithmic}[1]
        \STATE Use embedding model $\mathcal{E}$ to get item embeddings $\mathcal{E}(\mathcal{I})$ 
        \STATE Use $\mathcal{E}(\mathcal{I})$ and true user history 
        to train perturbation model $\mathcal{P}$
        \FOR{each user $u$}
            \FOR{$i$ from $1$ to $m$}
                \STATE $\tilde{\mathcal{H}}^u_i \leftarrow \mathcal{P}(\mathcal{H}^{u})$; $\tilde{Y}^u_i\leftarrow\mathcal{F}(\tilde{\mathcal{H}}^u_i)$
            \ENDFOR
            \STATE Construct perturbed input-output pairs $\{(\tilde{\mathcal{H}}^u_i,\tilde{Y}^u_i)\}_{i=1}^m$
            \STATE 
            $\theta_{\tilde{H}^u_{ij},\tilde{Y}^u_{i}} \leftarrow \mathcal{M}\Big(\{(\tilde{\mathcal{H}}^u_i,\tilde{Y}^u_i)\}_{i=1}^m \cup (\mathcal{H}^u,Y^u)\Big)$ 
            \STATE Rank $\theta_{\tilde{H}^u_{ij},Y^u}$ and select top-$k$ pairs $\{(H_j,Y^u)\}_{j=1}^k$ 
            \IF{$\exists H_{\min\{j\}}\in \mathcal{H}^{u}$}
                \STATE Generate causal explanation $H_{\min\{j\}}\Rightarrow Y^u$
            \ELSE 
                \STATE No explanation for the recommended item $Y^u$
            \ENDIF
        \ENDFOR
        \RETURN all causal explanations $H\Rightarrow Y^{u}$
    \end{algorithmic}
    \label{alg:model}
\end{algorithm}

\subsection{Causal Model for Post-Hoc Explanation}

We thus introduce our causal explanation framework for recommendation. 
Inspired by \cite{alvarez2017causal}, we divide our framework into two models: a perturbation model and a causal rule mining model. 
The overview of the model framework is shown in Fig.\ref{fig:model}. 
Before introducing our framework in detail, we define an important concept:
\begin{mydef}
(Causal Dependency): For a given pair of causal relation candidate $(H,Y^u)$, the causal dependency of the pair is the likelihood of the pair being a true causal relation.
\end{mydef}


\subsubsection{\textbf{Perturbation Model}}
\label{sec:perturbation-model}


To capture the causal dependency between items in history and the recommended items, we want to know what would happen if the user history had been different.
To avoid unknown influences caused by the length of input sequence (i.e., user history), we keep the input length unchanged, and only replace items in the sequence to create different sequences.
Ideally, for each item $H_j^u$ in a user's history $\mathcal{H}^u$, it will be replaced by all possible items in $\mathcal{I}$ to fully explore the influence that $H_j^u$ makes in the history.
However, the number of possible combinations will become impractical for the learning system, since recommender systems usually deal with hundreds of thousands or even millions of items. 
Therefore, we pursue a perturbation-based method that samples the sequences, which randomly replaces items in the original user history $\mathcal{H}^u$. 


There are various ways to obtain the perturbed version of user history, as long as they are similar to the original history.
The simplest solution is randomly selecting an item in $\mathcal{H}^u$ and replacing it with a randomly selected item from $\mathcal{I}$.
However, user histories are far from random selections.
Instead, a natural assumption is that the user behavior follows a certain distribution.
Here, we adopt VAE to learn such a distribution.
As shown in Figure \ref{fig:model}, we design a VAE-based perturbation method, which creates item sequences that are similar to but slightly different from a user's true behavior sequence, by sampling from a distribution in latent embedding space centered around the user's true sequence.


In detail, the VAE component consists of a probabilistic encoder $(\mu,\sigma) = \mathrm{ENC}(\mathcal{X})$ and a decoder $\tilde{\mathcal{X}} = \mathrm{DEC}(z)$. 
The encoder encodes a sequence of item embeddings $\mathcal{X}$ into latent embedding space, and extracts the variational information for the sequence, i.e., mean and variance of the latent embeddings under independent Gaussian distribution. 
The decoder generates a sequence of item embeddings $\tilde{\mathcal{X}}$ given a latent embedding $z$ sampled from the Gaussian distribution. 
Here, both $\mathcal{X}$ and $\tilde{\mathcal{X}}$ are ordered concatenations of pre-trained item embeddings based on pair-wise matrix factorization (BPRMF) \cite{rendle2012bpr}.
We follow the standard training regime of VAE by maximizing the variational lower bound of the data likelihood \cite{kingma2013auto}.
Specifically, the reconstruction error involved in this lower bound is calculated by a softmax across all items for each position of the input sequence. 
We observe that VAE can reconstruct the original data set accurately, while offering the power of perturbation.

After pretraining  $\mathrm{ENC}(\cdot)$ and $\mathrm{DEC}(\cdot)$, the variational nature of this model allow us to obtain perturbation $\tilde{\mathcal{H}}$ for any history $\mathcal{H}$. 
More specifically, it first extracts the mean and variance of the encoded item sequences in the latent space, and the perturbation model samples $m$ latent embeddings $z$ based on the above variational information.
These sampled embeddings $z$ are then passed to the decoder $\mathrm{DEC}(\cdot)$ to obtain the perturbed versions $\tilde{\mathcal{X}}$.
For now, each item embedding in $\tilde{\mathcal{X}}$ may not represent an actual item since it is a sampled vector from the latent space, as a result, we find its nearest neighbor in the candidate item set $\mathcal{I}$ through dot product similarity as the actual item. In this way, $\tilde{\mathcal{X}}$ is transformed into the final perturbed history $\tilde{\mathcal{H}}$.
One should keep in mind that the variance should be kept small during sampling, so that the resulting sequences are similar to the original sequence.


Finally, the generated perturbed data $\tilde{\mathcal{H}}$ together with the original $\mathcal{H}$ will be injected into the black-box recommendation model $\mathcal{F}$ to obtain the recommendation results $\tilde{Y}$ and $Y$, correspondingly. 
After completing this process, we will have $m$ perturbed input-output pairs: $\{(\tilde{\mathcal{H}}^u_i,\tilde{Y}^u_i)\}_{i=1}^m$, as well as the original pair $(\mathcal{H}^u,Y^u)$.

\subsubsection{\textbf{Causal Rule Learning Model}}
\label{sec:causal_rule_learning}
Denote $\mathcal{D}^u$ as the combined records of perturbed input-output pairs $\{(\tilde{\mathcal{H}}^u_i,\tilde{Y}^u_i)\}_{i=1}^m$ and the original pair $(\mathcal{H}^u,Y^u)$ for user $u$.
We aim to develop a causal model that first extracts causal dependencies between input and output items appeared in $\mathcal{D}^u$, and then selects the causal rule based on these inferred causal dependencies.

Let $\tilde{\mathcal{H}}^u_i=[\tilde{H}^u_{i1}, \tilde{H}^u_{i2}, \cdots, \tilde{H}^u_{in}]$ be the input sequence of the $i$-th record of $\mathcal{D}^u$. 
Let $\tilde{Y}^u_i$ represent the corresponding output. 
Note that this includes the original pair $(\mathcal{H}^u,Y^u)$.
The model should be able to infer the causal dependency $\theta_{\tilde{H}^u_{ij},\tilde{Y}^u_{i}}$ between input item $\tilde{H}^u_{ij}$ and output item $\tilde{Y}^u_{i}$.
We consider that the occurrence of a single output can be modeled as a logistic regression model on causal dependencies from all the input items in the sequence:
\begin{equation}
    P(\tilde{Y}^u_{i}|\tilde{\mathcal{H}}^u_{i}) = \sigma\Big(\sum_{j=1}^n \theta_{\tilde{H}^u_{ij},\tilde{Y}^u_{i}}\cdot\gamma^{n-j}\Big)\label{con:eq1}
\vspace{-5pt}
\end{equation}
where $\sigma$ is the sigmoid function defined as $\sigma(x) = (1+\exp(-x))^{-1}$ in order to scale the score to $[0,1]$.
Additionally, in recommendation task, the order of a user's previously interacted items may affect their causal dependency with the user's next interaction. 
A closer behavior tends to have a stronger effect on user's future behaviors, and behaviors are discounted if they happened earlier~\cite{hidasi2015session}. 
Therefore, we involve a weight decay parameter $\gamma$ to represent the time effect. Here $\gamma$ is a positive value less than one.

For an input-output pair in $\mathcal{D}^u$, the probability of its occurrence generated by Eq.\eqref{con:eq1} should be close to one. As a result, we learn the causal dependencies $\theta$ by maximizing the probability over $\mathcal{D}^u$.
When optimizing $\theta$, they are always initialized as zero to allow for no causation between two items. 
By learning this regression model, we are able to gradually increase $\theta$ until they converge to the point where the data likelihood of $\mathcal{D}$ is maximized.

After gathering all the causal dependencies, we select the items that have high $\theta$ scores to build causal explanations.
This involves a three-step procedure.
\begin{itemize}[leftmargin=*]
\item[1.] We select those causal dependencies $\theta_{\tilde{H}^u_{ij},\tilde{Y}^u_{i}}$ whose output is the original output $Y^u$ (i.e., $\tilde{Y}^u_{i}=Y^u$). Note that these $(\tilde{H}^u_{ij},Y^u)$ pairs may come from either the original sequence or perturbed sequences, because when a perturbed sequence is fed into the black-box recommendation model, the output may happen to be the same as the original sequence $Y^u$. 
\item[2.] We sort the above selected causal dependencies in descending order and take the top-$k$ $(\tilde{H}^u_{ij},Y^u)$ pairs.
\item[3.] If there exist one or more pairs in these top-$k$ pairs, whose cause item $\tilde{H}^u_{ij}$ appears in the user's original input sequence $\mathcal{H}^u$, then we pick such pair of the highest rank, and construct $\tilde{H}^u_{ij}\Rightarrow Y^u$ as the causal explanation for the given user.
Otherwise, i.e., no cause item appears in the user history, then we output no causal explanation for the user. 
\end{itemize}

Note that the extracted causal explanation is personalized since the algorithm is applied on $\mathcal{D}^u$, which only contains records centered around the user's original record $(\mathcal{H}^u,Y^u)$, while collaborative learning among users is indirectly modeled by the VAE-based perturbation model. The overall algorithm is provided in Alg.\ref{alg:model}.



\section{Experiments}
In this part, we conduct experiments to show what causal relationships our model can capture and how they can serve as an intuitive explanation for the black-box recommendation model. 

\subsection{Dataset Description}
We evaluate our proposed causal explanation framework against baselines on two datasets. The first dataset is MovieLens100k\footnote{https://grouplens.org/datasets/movielens/}. 
This dataset consists of information about users, movies and ratings. In this dataset, each user has rated at least 20 movies, and each movie can belong to several genres. The second dataset is the office product dataset from Amazon\footnote{https://nijianmo.github.io/amazon/}, which contains the user-item interactions from May 1996 to July 2014. The original dataset is 5-core. To achieve sequential recommendation with input length of 5, we select the users with at least 15 purchases and the items with at least 10 interactions.

Since our framework is used to explain sequential recommendation models, we split the dataset chronologically. 
To learn the pre-trained item embeddings based on BPRMF \cite{rendle2012bpr} (section \ref{sec:perturbation-model}), we take the last 6 interactions from each user to construct the testing set, and use all previous interactions from each user as the training set. 
To avoid data leakage, when testing the black-box recommendation models and our VAE-based perturbation model, we only use the last 6 interactions from each user (i.e., the testing set of the pre-training stage). 
Following common practice, 
we adopt the leave-one-out protocol, i.e., among the 6 interactions, we use the last one for testing, and the previous 5 as input to recommendation models.
A brief summary of the data is shown in Table \ref{tab:data}. 

\begin{table}[t]
    \centering
    \setlength{\tabcolsep}{1pt}
    \begin{tabular}{ccccccc}
        \toprule
        Dataset & \# users & \# items & \# interactions & \# train & \# test & sparsity \\\midrule
        Movielens & 943 & 1682 & 100,000 & 95,285 & 14,715 & 6.3\%\\
        Amazon & 573 & 478 & 13,062 & 9,624 & 3,438 & 4.7\%\\\bottomrule
    \end{tabular}
    \caption{Summary of the Datasets}
    \vspace{-10pt}
    \label{tab:data}
    \vspace{-10pt}
\end{table}

\begin{table*}[t]
    \centering
    \setlength{\tabcolsep}{4pt}
    \begin{tabular}{ccccccccccccc}
        \toprule
        Dataset & \multicolumn{6}{c}{Movielens 100k} & \multicolumn{6}{c}{Amazon}\\\cmidrule(lr){1-1}\cmidrule(lr){2-7}\cmidrule(lr){8-13}
        Method & \multicolumn{3}{c}{Causal} & \multicolumn{3}{c}{Association} & \multicolumn{3}{c}{Causal} & \multicolumn{3}{c}{Association}\\\cmidrule(lr){2-4}\cmidrule(lr){5-7}\cmidrule(lr){8-10}\cmidrule(lr){11-13}
        Parameter & $k=1$ & $k=2$ & $k=3$ & Support & Confidence & Lift & $k=1$ & $k=2$ & $k=3$ & Support & Confidence & Lift\\\midrule
        FPMC & 96.50\% & 99.25\% & 99.57\% & 14.32\% & 13.47\% & 13.47\%& 95.11\% & 98.95\% & 99.82\% & 12.57\% & 12.04\% & 12.04\%\\
        GRU4Rec & 98.51\% & 99.36\% & 99.68\% & 7.423\% & 7.635\% & 7.635\% & 95.94\% & 99.48\% & 99.65\% & 8.202\% & 8.551\% & 8.551\%\\
        Caser & 97.03\% & 99.15\% & 99.57\% & 8.801\% & 8.165\% & 8.165\% & 95.99\% & 99.30\% & 99.65\% & 9.250\% & 8.901\% & 8.901\%\\
        \bottomrule
    \end{tabular}
    \caption{Results of Model Fidelity. Our causal explanation framework is tested under the number of candidate causal explanations $k=1,2,3$. The association explanation framework is tested under support, confidence, and lift thresholds, respectively.}
    \vspace{-15pt}
    \label{tab:fidelity}
\end{table*}

\subsection{Experimental Settings}
\label{sec:experimental_setting}

We adopt the following methods to train black-box sequential recommendation models and to extract traditional association rules as comparative explanations. We include both shallow and deep models for experiment.

\textbf{FPMC}~\cite{rendle2010factorizing}: The Factorized Personalized Markov Chain model, which combines matrix factorization and Markov chains to capture user's personalized sequential behavior patterns for prediction\footnote{https://github.com/khesui/FPMC}.

\textbf{GRU4Rec}~\cite{hidasi2015session}: A session-based recommendation model, which uses recurrent neural networks -- in particular, Gated Recurrent Units (GRU) -- to capture sequential patterns for prediction\footnote{https://github.com/hungthanhpham94/GRU4REC-pytorch}.

\textbf{Caser}~\cite{tang2018personalized}: The ConvolutionAl Sequence Embedding Recommendation (Caser) model, which adopts convolutional filters over recent items to learn the sequential patterns for prediction\footnote{https://github.com/graytowne/caser\_pytorch}.

\textbf{Association Rule}~\cite{peake2018explanation}: A post-hoc explanation model, which learns the item-item association rules as item-level explanations\footnote{https://pypi.org/project/apyori/}.

For black-box recommendation models FPMC, GRU4Rec and Caser, we adopt their best parameter selection in their corresponding public implementation. For the association rule-based explanation model, we follow the recommendations in \cite{peake2018explanation} to set the optimal parameters: support = 0.1, confidence = 0.1, lift = 0.1, length = 2 for \textit{MovieLens100k}, and support = 0.01, confidence = 0.01, lift = 0.01, length = 2 for \textit{Amazon} dataset due to its smaller scale.

For our causal rule learning framework, we set the item embedding size as 16, both the VAE encoder and decoder are Multi-Layer Perceptrons (MLP) with two hidden layers, and each layer consists of 1024 neurons. The default number of perturbed input-output pairs is $m=500$ on both datasets. The default time decay factor is $\gamma=0.7$. We will discuss the influence of perturbation times $m$ and time decay factor $\gamma$ in the experiments. 

In the following, we will apply both association rule learning and causal rule learning frameworks on the black-box recommendation models to evaluate and compare the association explanations and causal explanations. In particular, we evaluate our framework from two perspectives. First, we verify that the causal rules learned by our framework represent highly probable causal relationships (explanation quality). Second, we show that our model has the ability to offer explanations for most recommendations (explanation fidelity). Additionally, we shed light on how our model differs from other models on statistical metrics.


\subsection{Causality Verification}
We first verify the quality of the extracted causal rules. Here, we adopt the following widely used definition of causation, which is introduced by Pearl \cite{pearl2000causality}:
\begin{equation}\label{eq:verification}
     \Pr(\text{effect}|do(\text{cause})) \textgreater \Pr(\text{effect}|do(\neg\text{cause}))
\end{equation}
where $do(c)$ represents an external intervention, which compels the truth of $c$, and $do(\neg c)$ compels the truth of not $c$. Actually, the conditional probability $\Pr(e|c)$ represents a probability resulting from a passive observation of $c$, which rarely coincides with $\Pr(e|do(c))$. We evaluate how many percentage of our extracted causal explanations really satisfy Eq.\eqref{eq:verification}.

Suppose the perturbation model (section \ref{sec:perturbation-model}) creates $m$ perturbed input-output pairs for each user $u$: $\{(\tilde{\mathcal{H}}^u_i,\tilde{Y}^u_i)\}_{i=1}^m$, plus the original pair $(\mathcal{H}^u,Y^u)$. Here $\tilde{\mathcal{H}}^u$ is created by our perturbation model (i.e. not observed in the original data), and thus observing $\tilde{\mathcal{H}}^u$ implies we have $do(\tilde{\mathcal{H}}^u)$ in advance. Let $H\Rightarrow Y^u$ be the causal explanation extracted by the casual rule learning model (section \ref{sec:causal_rule_learning}). Then we estimate the probability based on these $m+1$ total pairs as,
\begin{equation}
\small
\begin{aligned}
    &\Pr(\text{effect}|do(\text{cause}))=\Pr(Y^u|H)=\frac{\#\text{Pairs}(H\in\tilde{\mathcal{H}}^u\land Y=Y^u)}{\#\text{Pairs}(H\in\tilde{\mathcal{H}}^u)}\\
    &\Pr(\text{effect}|do(\neg\text{cause}))=\Pr(Y^u|\neg H)=\frac{\#\text{Pairs}(H\notin\tilde{\mathcal{H}}^u\land Y=Y^u)}{\#\text{Pairs}(H\notin\tilde{\mathcal{H}}^u)}
\end{aligned}
\end{equation}
and $H\Rightarrow Y^u$ is considered as a reliable causal rule if Eq.\eqref{eq:verification} is satisfied. The intuition here is that a causal rule should guarantee that the probability of seeing $Y^u$ when $H$ is purchased should be higher than the probability of seeing $Y^u$ when $H$ is not purchased.



We calculate the percentage of our extracted causal explanations that satisfy Eq.\eqref{eq:verification}. We tune the parameter $k$ from 1 to 3, where $k$ is the number of candidate causal explanations we consider for each user in the causal rule learning model (section \ref{sec:causal_rule_learning} step 2 and 3).
The results of causality verification on two datasets and three recommendation models are shown in Table \ref{tab:verification}.

Based on the results we can see that in most cases 90\% or more of the extracted causal explanations are reliable. We also see that with the increasing of $k$, the percentage tends to decease for all three recommendation models on both datasets. Intuitively, this means that unreliable causal relation candidates may be introduced if too many candidate pairs are considered in the causal rule learning model, which is reasonable. However, if too few candidates are considered (i.e., $k$ is too small), many users may not receive explanations at all (i.e., model fidelity will decrease). This will be further analyzed in the next subsection.


\begin{table}[t]
    \centering
    \setlength{\tabcolsep}{3pt}
    \begin{tabular}{ccccccc}
        \toprule
        Dataset & \multicolumn{3}{c}{Movielens} & \multicolumn{3}{c}{Amazon}\\\cmidrule(lr){2-4}\cmidrule(lr){5-7}
        $k$ & 1 & 2 & 3 & 1 & 2 & 3\\\midrule
        FPMC & 94.83\% & 94.23\% & 93.92\% & 91.56\% & 90.48\% & 90.03\%\\
        GRU4Rec & 97.84\% & 97.76\% & 97.75\% & 94.91\% & 94.20\% & 93.87\%\\
        Caser & 97.15\% & 96.89\% & 96.69\% & 98.91\% & 98.25\% & 98.14\%\\\bottomrule
    \end{tabular}
    \caption{The percentage of reliable causal explanations that satisfy the inequality in Eq.\eqref{eq:verification}.}
    \vspace{-27pt}
    \label{tab:verification}
\end{table}

\subsection{Model Fidelity}
An important evaluation measure for explanation models is model fidelity, i.e., how many percentage of the recommendation results can be explained by the model \cite{zhang2018explainable}. 
The results of model fidelity are shown in Table \ref{tab:fidelity}. In this experiment, we still tune the number of candidate causal explanations $k$ from 1 to 3. For the association rule explanation model (section \ref{sec:experimental_setting}), we test three versions of the association rule learning algorithm as introduced in \cite{peake2018explanation}, i.e., the association rules are filtered by support, confidence, and lift thresholds, respectively. 


We can see that on both datasets, our causal explanation framework can generate explanations for almost all of the recommended items, while the association explanation approach can only provide explanations for significantly fewer recommendations. The underlying reason is that association explanations have to be extracted based on the original input-output pairs, which limits the number of pairs that we can use for rule extraction. However, based on the perturbation model, our causal explanation framework is able to create many counterfactual examples to assist causal rule learning, which makes it possible to go beyond the limited original data to extract causal explanations.

Another interesting observation is that GRU4Rec and Caser have significantly ($p<0.01$) lower fidelity than FPMC when explained by the association model. This is reasonable because FPMC is a Markov-based model that directly learns the correlation of adjacent items in a sequence, as a result, it is easier to extract association rules between inputs and outputs for the model. However, it also means that the fidelity performance of the association approach highly depends on the recommendation model being explained. Meanwhile, we see that our causal approach achieves comparably good fidelity on all three recommendation models, because the perturbation model is able to create sufficiently many counterfactual examples to break the correlation of frequently co-occurring items in the input sequence. This indicates the robustness of our causal explanation framework in terms of model fidelity.



\subsection{Influence of Parameters}
In this section, we discuss the influence of two important parameters. The first one is time decay parameter $\gamma$ -- in our framework, when explaining the sequential recommendation models, earlier interactions in the sequence will have discounted effects to the recommended item. A proper time decay parameter helps the framework to reduce noise signals when learning patterns from the sequence. The second parameter is the number of perturbed input-output pairs $m$ -- in our framework, we use perturbations to create counterfactual examples for causal learning, but there may exist trade-off between efficiency and performance. We will analyze the influence of these two parameters.

\begin{figure}[t]
\captionsetup[sub]{font=small,labelfont=normalfont,textfont=normalfont}
\centering
\hspace{-15pt}
\begin{subfigure}{0.24\textwidth}
\label{fig:GRU4Rec_ml100k_time_decay}
        \includegraphics[scale=0.3]{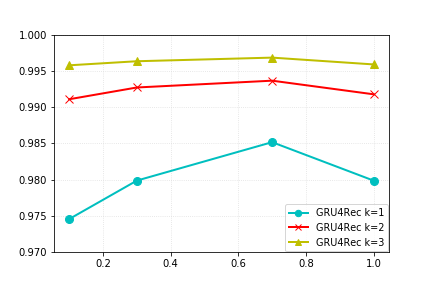}
        \vspace{-20pt}
        \caption{GRU4Rec on Movielens}
\end{subfigure}
\begin{subfigure}{0.24\textwidth}
\label{fig:GRU4Rec_amazon_time_decay}
        \includegraphics[scale=0.3]{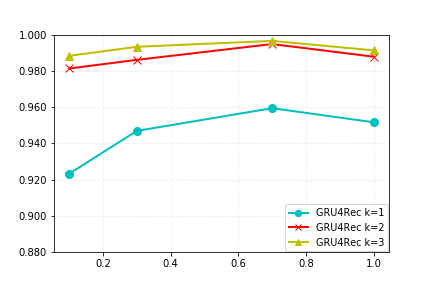}
        \vspace{-20pt}
        \caption{GRU4Rec on Amazon}
\end{subfigure}
\medskip
\hspace{-15pt}
\begin{subfigure}{0.24\textwidth}
\label{fig:k3_ml100k_time_decay}
        \includegraphics[scale=0.3]{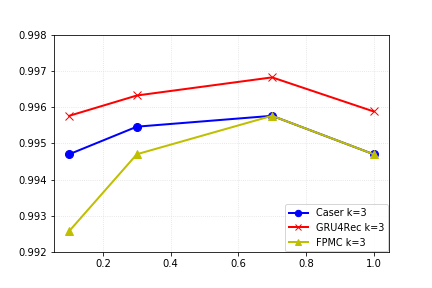}
        \vspace{-20pt}
        \caption{$k=3$ on Movielens}
\end{subfigure}
\begin{subfigure}{0.24\textwidth}
\label{fig:k3_amazon_time_decay}
        \includegraphics[scale=0.3]{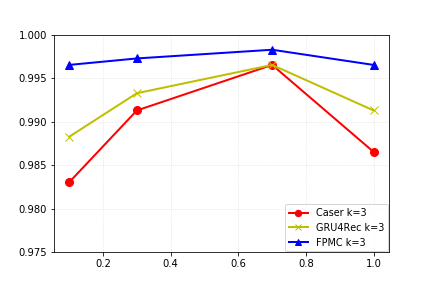}
        \vspace{-20pt}
        \caption{$k=3$ on Amazon}
\end{subfigure}
\vspace{-10pt}
\caption{Model fidelity on different time decay parameters $\gamma$. $x$-axis is the time decay parameter $\gamma\in\{0.1, 0.3, 0.7, 1\}$ and $y$-axis is the model fidelity. 
The upper two sub-figures plot the model fidelity for GRU4Rec on different number of candidate causal explanations $k=1,2,3$, and the lower two sub-figures plot the model fidelity for different recommendation models when fixing $k=3$.}
\label{fig:time}
\vspace{-10pt}
\end{figure}

\textbf{Time Decay Effect}: 
Figure \ref{fig:time} shows the influence of $\gamma$ on different recommendation models and datasets. From the result we can see that the time decay effect $\gamma$ indeed affects the model performance on fidelity. In particular, when $\gamma$ is small, the previous interactions in a sequence are more likely to be ignored, which thus reduces the performance on model fidelity. When $\gamma$ is large (e.g., $\gamma=1$), old interactions will have equal importance with latest interactions, which also hurts the performance. We can see from the results that the best performance is achieved at about $\gamma=0.7$ on both datasets.

\begin{figure}[t]
\captionsetup[sub]{font=small,labelfont=normalfont,textfont=normalfont}
\centering
\hspace{-15pt}
\begin{subfigure}{0.24\textwidth}
\label{fig:GRU4Rec_ml100k_time_decay}
        \includegraphics[scale=0.3]{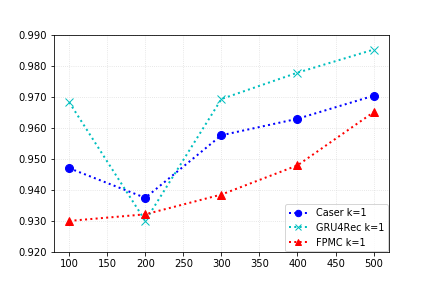}
        \vspace{-20pt}
        \caption{Model Fidelity on Movielens}
\end{subfigure}
\begin{subfigure}{0.24\textwidth}
\label{fig:GRU4Rec_amazon_time_decay}
        \includegraphics[scale=0.3]{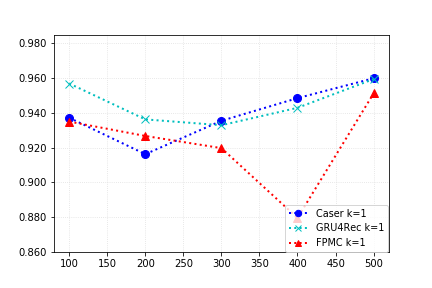}
        \vspace{-20pt}
        \caption{Model Fidelity on Amazon}
\end{subfigure}
\medskip
\hspace{-15pt}
\begin{subfigure}{0.24\textwidth}
\label{fig:k3_ml100k_time_decay}
        \includegraphics[scale=0.3]{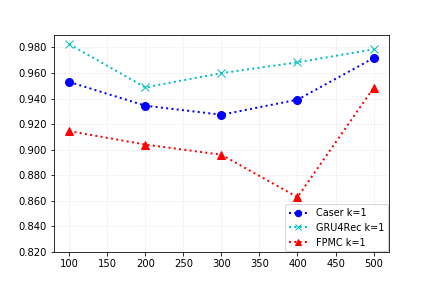}
        \vspace{-20pt}
        \caption{Percentage on Movielens}
\end{subfigure}
\begin{subfigure}{0.24\textwidth}
\label{fig:k3_amazon_time_decay}
        \includegraphics[scale=0.3]{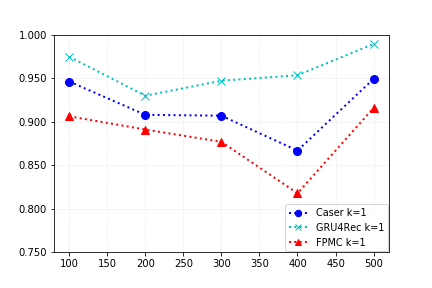}
        \vspace{-20pt}
        \caption{Percentage on Amazon}
\end{subfigure}
\vspace{-10pt}
\caption{Model fidelity and the percentage of verified causal explanations (Eq.\eqref{eq:verification}) on different number of perturbed pairs $m$. $x$-axis is the number of perturbed pairs $m$. $y$-axis is model fidelity. The upper two sub-figures plot the model fidelity for different recommendation model on $k=1$, and the lower two sub-figures plot the percentage of verified causal explanations for different recommendation model on $k=1$.}
\label{fig:pert_time}
\vspace{-10pt}
\end{figure}

\textbf{Number of Perturbations}: 
Figure \ref{fig:pert_time} shows the influence for the number of perturbed input-output pairs $m$.
A basic observation from Figure \ref{fig:pert_time} is that when $m$ increases, both model fidelity and the percentage of verified rules will decrease first and then increase. The underlying reason is as follows. 

When $m$ is small, the variance of the perturbed input-output pairs will be small, and thus almost any change in the input sequence will be determined as a cause. For example, suppose the original input-output pair is $A, B, C \rightarrow Y$. In the extreme case where $m=1$, we will have only one perturbed pair, e.g., $A, \tilde{B}, C \rightarrow \tilde{Y}$. According to the causal rule learning model (section \ref{sec:causal_rule_learning}), if $\tilde{Y}\neq Y$, then $B\Rightarrow Y$ will be the causal explanation since the change of $B$ results in a different output, while if $\tilde{Y}=Y$, then either $A\Rightarrow Y$ or $C\Rightarrow Y$ will be the causal explanation since their $\theta$ scores will be higher than $B$ or $\tilde{B}$. In either case, the model fidelity and percentage of verified causal rules will be $100\%$. However, in this case, the results do not present statistical meanings since they are estimated on a very small amount of examples.

When $m$ increases but not large enough, then random noise examples created by the perturbation model will reduce the model fidelity. Still consider the above example, if many pairs with the same output $Y$ are created, then the model may find other items beyond $A,B,C$ as the cause, which will result in no explanation for the original sequence. However, if we continue to increase $m$ to sufficiently large numbers, such noise will be statistically offset, and thus the model fidelity and percentages will increase again. In the most ideal case, we would create all of the $|\mathcal{H}|^{|\mathcal{I}|}$ sequences for causal rule learning, where $|\mathcal{H}|$ is the number of item slots in the input sequence, and $|\mathcal{I}|$ is the total number of items in the dataset. However, $|\mathcal{H}|^{|\mathcal{I}|}$ would be a huge number that makes it computational infeasible for causal rule learning. In practice, we only need to specify $m$ sufficiently large. Based on Chebyshev's Inequality, we find that $m=500$ already gives >95\% confidence that the estimated probability error is <0.1.


\subsection{Case Study}
\label{sec:explanation}
\begin{figure}[t!]
    \centering
    \includegraphics[width=\linewidth]{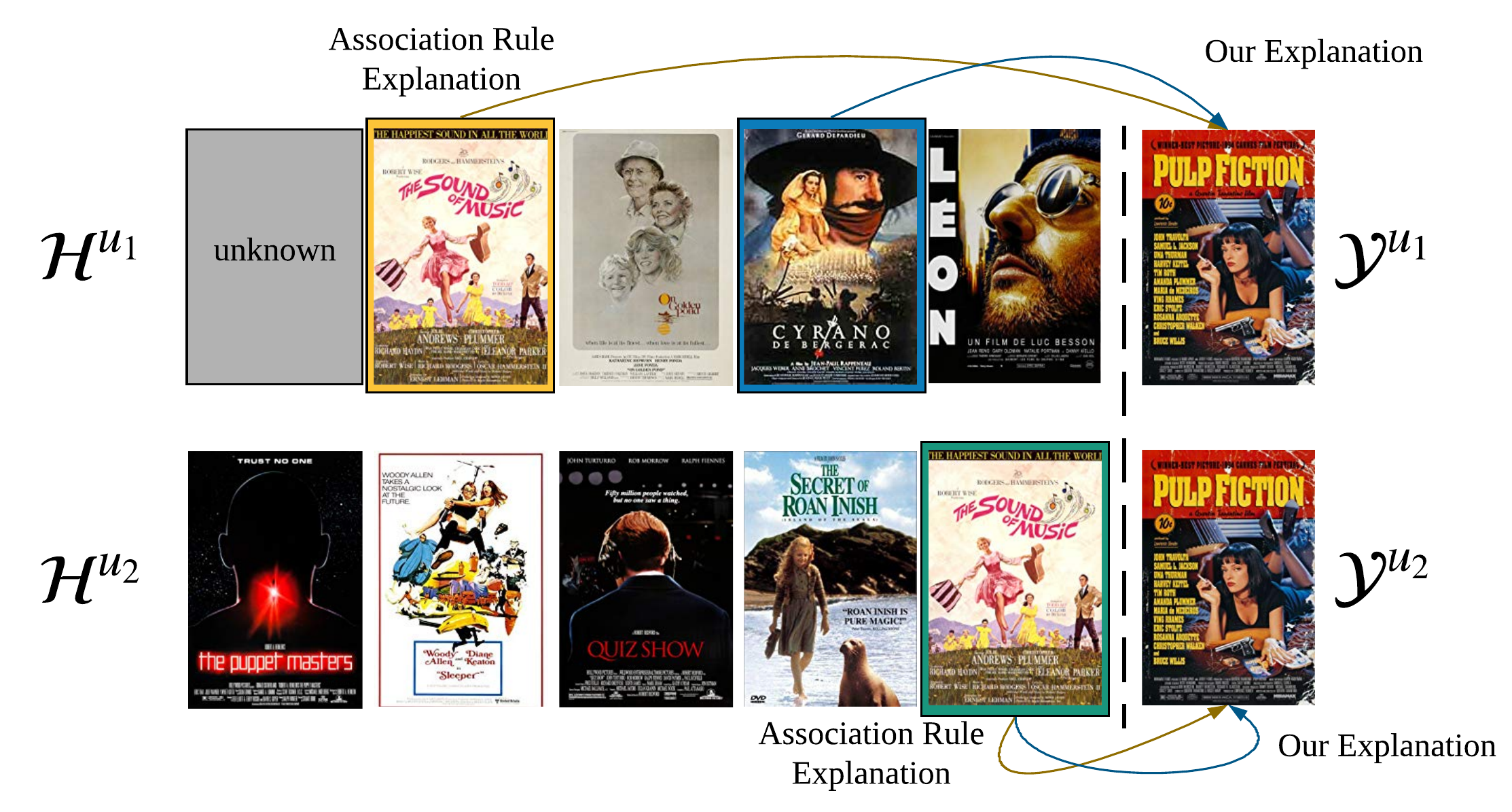}
    \vspace{-25pt}
    \caption{
    A case study on MovieLens by the Caser model. The first movie for $u_1$ is unknown in the dataset.
    }
    \vspace{-10pt}
    \label{fig:case_study}
\end{figure}

In this section, we provide a simple qualitative case study to compare causal explanations and association explanations. Compared with the association explanation model, our model is capable of generating personalized explanations, which means that even if the recommendation model recommends the same item for two different users and the users have overlapped histories, our model still has the potential to generate different explanations for different users. However, the association model will provide the same explanation since the association rules are extracted based on global records.
An example by the Caser recommendation model on \textit{MovieLens100k} dataset is shown in Figure \ref{fig:case_study},
where two users with one commonly watched movie (\textit{The Sound of Music}) get exactly same recommendation (\textit{Pulp Fiction}). The association model provides the overlapped movie as an explanation for the two different users, while our model can generate personalized explanation for different users even when they got the same recommendation. 
\vspace{-5pt}
\section{Conclusions}
Recommender systems are widely used in our daily life. 
Effective recommendation mechanisms usually work through black-box models, resulting in the lack of transparency.
In this paper, we extract causal rules from user history to provide personalized, item-level, post-hoc explanations for the black-box sequential recommendation models.
The causal explanations are extracted through a perturbation model and a causal rule learning model. 
We conduct experiments on real-world datasets, and apply our explanation framework to several state-of-the-art sequential recommendation models.
Experimental results verified the quality and fidelity of the causal explanations extracted by our framework.

In this work, we only considered item-level causal relationships, while in the future, it would be interesting to explore causal relations on feature-level external data such as textual user reviews, which can help to generate finer-grained causal explanations. 

\section*{Acknowledgement}
This work was supported in part by NSF IIS-1910154. Any opinions, findings, conclusions or recommendations expressed in this material are those of the authors and do not necessarily reflect those of the sponsors.

\vspace{-5pt}
\balance
\bibliographystyle{ACM-Reference-Format}
\bibliography{sample-base}


\begin{thebibliography}{36}


\ifx \showCODEN    \undefined \def \showCODEN     #1{\unskip}     \fi
\ifx \showDOI      \undefined \def \showDOI       #1{#1}\fi
\ifx \showISBNx    \undefined \def \showISBNx     #1{\unskip}     \fi
\ifx \showISBNxiii \undefined \def \showISBNxiii  #1{\unskip}     \fi
\ifx \showISSN     \undefined \def \showISSN      #1{\unskip}     \fi
\ifx \showLCCN     \undefined \def \showLCCN      #1{\unskip}     \fi
\ifx \shownote     \undefined \def \shownote      #1{#1}          \fi
\ifx \showarticletitle \undefined \def \showarticletitle #1{#1}   \fi
\ifx \showURL      \undefined \def \showURL       {\relax}        \fi
\providecommand\bibfield[2]{#2}
\providecommand\bibinfo[2]{#2}
\providecommand\natexlab[1]{#1}
\providecommand\showeprint[2][]{arXiv:#2}

\bibitem[\protect\citeauthoryear{Ai, Azizi, Chen, and Zhang}{Ai
  et~al\mbox{.}}{2018}]%
        {ai2018learning}
\bibfield{author}{\bibinfo{person}{Qingyao Ai}, \bibinfo{person}{Vahid Azizi},
  \bibinfo{person}{Xu Chen}, {and} \bibinfo{person}{Yongfeng Zhang}.}
  \bibinfo{year}{2018}\natexlab{}.
\newblock \showarticletitle{Learning heterogeneous knowledge base embeddings
  for explainable recommendation}.
\newblock \bibinfo{journal}{\emph{Algorithms}} \bibinfo{volume}{11},
  \bibinfo{number}{9} (\bibinfo{year}{2018}), \bibinfo{pages}{137}.
\newblock


\bibitem[\protect\citeauthoryear{Alvarez-Melis and Jaakkola}{Alvarez-Melis and
  Jaakkola}{2017}]%
        {alvarez2017causal}
\bibfield{author}{\bibinfo{person}{David Alvarez-Melis} {and}
  \bibinfo{person}{Tommi~S Jaakkola}.} \bibinfo{year}{2017}\natexlab{}.
\newblock \showarticletitle{A causal framework for explaining the predictions
  of black-box sequence-to-sequence models}.
\newblock \bibinfo{journal}{\emph{Proceedings of the 2017 Conference on
  Empirical Methods in Natural Language Processing}} (\bibinfo{year}{2017}).
\newblock


\bibitem[\protect\citeauthoryear{Bonner and Vasile}{Bonner and Vasile}{2018}]%
        {bonner2018causal}
\bibfield{author}{\bibinfo{person}{Stephen Bonner} {and}
  \bibinfo{person}{Flavian Vasile}.} \bibinfo{year}{2018}\natexlab{}.
\newblock \showarticletitle{Causal embeddings for recommendation}. In
  \bibinfo{booktitle}{\emph{Proceedings of the 12th ACM Conference on
  Recommender Systems}}. ACM.
\newblock


\bibitem[\protect\citeauthoryear{Buesing, Weber, Zwols, Racaniere, Guez,
  Lespiau, and Heess}{Buesing et~al\mbox{.}}{2019}]%
        {buesing2018woulda}
\bibfield{author}{\bibinfo{person}{Lars Buesing}, \bibinfo{person}{Theophane
  Weber}, \bibinfo{person}{Yori Zwols}, \bibinfo{person}{Sebastien Racaniere},
  \bibinfo{person}{Arthur Guez}, \bibinfo{person}{Jean-Baptiste Lespiau}, {and}
  \bibinfo{person}{Nicolas Heess}.} \bibinfo{year}{2019}\natexlab{}.
\newblock \showarticletitle{Woulda, Coulda, Shoulda: Counterfactually-Guided
  Policy Search}. In \bibinfo{booktitle}{\emph{International Conference on
  Learning Representations}}.
\newblock


\bibitem[\protect\citeauthoryear{Chen, Zhuang, Hong, Ao, Xie, and He}{Chen
  et~al\mbox{.}}{2018b}]%
        {chen2018attention}
\bibfield{author}{\bibinfo{person}{Jingwu Chen}, \bibinfo{person}{Fuzhen
  Zhuang}, \bibinfo{person}{Xin Hong}, \bibinfo{person}{Xiang Ao},
  \bibinfo{person}{Xing Xie}, {and} \bibinfo{person}{Qing He}.}
  \bibinfo{year}{2018}\natexlab{b}.
\newblock \showarticletitle{Attention-driven factor model for explainable
  personalized recommendation}. In \bibinfo{booktitle}{\emph{The 41st
  International ACM SIGIR Conference on Research \& Development in Information
  Retrieval}}. ACM, \bibinfo{pages}{909--912}.
\newblock


\bibitem[\protect\citeauthoryear{Chen, Xu, Zhang, Tang, Cao, Qin, and Zha}{Chen
  et~al\mbox{.}}{2018a}]%
        {chen2018sequential}
\bibfield{author}{\bibinfo{person}{Xu Chen}, \bibinfo{person}{Hongteng Xu},
  \bibinfo{person}{Yongfeng Zhang}, \bibinfo{person}{Jiaxi Tang},
  \bibinfo{person}{Yixin Cao}, \bibinfo{person}{Zheng Qin}, {and}
  \bibinfo{person}{Hongyuan Zha}.} \bibinfo{year}{2018}\natexlab{a}.
\newblock \showarticletitle{Sequential recommendation with user memory
  networks}. In \bibinfo{booktitle}{\emph{Proceedings of the eleventh ACM
  international conference on web search and data mining}}.
  \bibinfo{pages}{108--116}.
\newblock


\bibitem[\protect\citeauthoryear{Costa, Ouyang, Dolog, and Lawlor}{Costa
  et~al\mbox{.}}{2018}]%
        {costa2018automatic}
\bibfield{author}{\bibinfo{person}{Felipe Costa}, \bibinfo{person}{Sixun
  Ouyang}, \bibinfo{person}{Peter Dolog}, {and} \bibinfo{person}{Aonghus
  Lawlor}.} \bibinfo{year}{2018}\natexlab{}.
\newblock \showarticletitle{Automatic generation of natural language
  explanations}. In \bibinfo{booktitle}{\emph{Proceedings of the 23rd
  International Conference on Intelligent User Interfaces Companion}}. ACM,
  \bibinfo{pages}{57}.
\newblock


\bibitem[\protect\citeauthoryear{Fu, Xian, Gao, Zhao, Huang, Ge, Xu, Geng,
  Shah, Zhang, et~al\mbox{.}}{Fu et~al\mbox{.}}{2020}]%
        {fu2020fairness}
\bibfield{author}{\bibinfo{person}{Zuohui Fu}, \bibinfo{person}{Yikun Xian},
  \bibinfo{person}{Ruoyuan Gao}, \bibinfo{person}{Jieyu Zhao},
  \bibinfo{person}{Qiaoying Huang}, \bibinfo{person}{Yingqiang Ge},
  \bibinfo{person}{Shuyuan Xu}, \bibinfo{person}{Shijie Geng},
  \bibinfo{person}{Chirag Shah}, \bibinfo{person}{Yongfeng Zhang},
  {et~al\mbox{.}}} \bibinfo{year}{2020}\natexlab{}.
\newblock \showarticletitle{Fairness-Aware Explainable Recommendation over
  Knowledge Graphs}.
\newblock \bibinfo{journal}{\emph{SIGIR}} (\bibinfo{year}{2020}).
\newblock


\bibitem[\protect\citeauthoryear{He and McAuley}{He and McAuley}{2016}]%
        {he2016fusing}
\bibfield{author}{\bibinfo{person}{Ruining He} {and} \bibinfo{person}{Julian
  McAuley}.} \bibinfo{year}{2016}\natexlab{}.
\newblock \showarticletitle{Fusing similarity models with markov chains for
  sparse sequential recommendation}. In \bibinfo{booktitle}{\emph{2016 IEEE
  16th International Conference on Data Mining (ICDM)}}. IEEE,
  \bibinfo{pages}{191--200}.
\newblock


\bibitem[\protect\citeauthoryear{Hidasi, Karatzoglou, Baltrunas, and
  Tikk}{Hidasi et~al\mbox{.}}{2016}]%
        {hidasi2015session}
\bibfield{author}{\bibinfo{person}{Bal{\'a}zs Hidasi},
  \bibinfo{person}{Alexandros Karatzoglou}, \bibinfo{person}{Linas Baltrunas},
  {and} \bibinfo{person}{Domonkos Tikk}.} \bibinfo{year}{2016}\natexlab{}.
\newblock \showarticletitle{Session-based recommendations with recurrent neural
  networks}. In \bibinfo{booktitle}{\emph{International Conference on Learning
  Representations}}.
\newblock


\bibitem[\protect\citeauthoryear{Huang, Zhao, Dou, Wen, and Chang}{Huang
  et~al\mbox{.}}{2018}]%
        {huang2018improving}
\bibfield{author}{\bibinfo{person}{Jin Huang}, \bibinfo{person}{Wayne~Xin
  Zhao}, \bibinfo{person}{Hongjian Dou}, \bibinfo{person}{Ji-Rong Wen}, {and}
  \bibinfo{person}{Edward~Y Chang}.} \bibinfo{year}{2018}\natexlab{}.
\newblock \showarticletitle{Improving sequential recommendation with
  knowledge-enhanced memory networks}. In \bibinfo{booktitle}{\emph{The 41st
  International ACM SIGIR Conference on Research \& Development in Information
  Retrieval}}. ACM, \bibinfo{pages}{505--514}.
\newblock


\bibitem[\protect\citeauthoryear{Imbens and Rubin}{Imbens and Rubin}{2015}]%
        {imbens2015causal}
\bibfield{author}{\bibinfo{person}{Guido~W Imbens} {and}
  \bibinfo{person}{Donald~B Rubin}.} \bibinfo{year}{2015}\natexlab{}.
\newblock \bibinfo{booktitle}{\emph{Causal inference in statistics, social, and
  biomedical sciences}}.
\newblock \bibinfo{publisher}{Cambridge University Press}.
\newblock


\bibitem[\protect\citeauthoryear{Joachims, Swaminathan, and Schnabel}{Joachims
  et~al\mbox{.}}{2017}]%
        {joachims2017unbiased}
\bibfield{author}{\bibinfo{person}{Thorsten Joachims}, \bibinfo{person}{Adith
  Swaminathan}, {and} \bibinfo{person}{Tobias Schnabel}.}
  \bibinfo{year}{2017}\natexlab{}.
\newblock \showarticletitle{Unbiased learning-to-rank with biased feedback}. In
  \bibinfo{booktitle}{\emph{Proceedings of the Tenth ACM International
  Conference on Web Search and Data Mining}}. ACM, \bibinfo{pages}{781--789}.
\newblock


\bibitem[\protect\citeauthoryear{Kingma and Welling}{Kingma and
  Welling}{2014}]%
        {kingma2013auto}
\bibfield{author}{\bibinfo{person}{Diederik~P Kingma} {and}
  \bibinfo{person}{Max Welling}.} \bibinfo{year}{2014}\natexlab{}.
\newblock \showarticletitle{Auto-Encoding Variational Bayes}.
\newblock \bibinfo{journal}{\emph{Proceedings of the 2nd International
  Conference on Learning Representations (ICLR)}}.
\newblock


\bibitem[\protect\citeauthoryear{Li, Quan, Peng, Qi, Deng, and Wu}{Li
  et~al\mbox{.}}{2019}]%
        {li2019capsule}
\bibfield{author}{\bibinfo{person}{Chenliang Li}, \bibinfo{person}{Cong Quan},
  \bibinfo{person}{Li Peng}, \bibinfo{person}{Yunwei Qi},
  \bibinfo{person}{Yuming Deng}, {and} \bibinfo{person}{Libing Wu}.}
  \bibinfo{year}{2019}\natexlab{}.
\newblock \showarticletitle{A Capsule Network for Recommendation and Explaining
  What You Like and Dislike}. In \bibinfo{booktitle}{\emph{Proceedings of the
  42nd International ACM SIGIR Conference on Research and Development in
  Information Retrieval}}. ACM, \bibinfo{pages}{275--284}.
\newblock


\bibitem[\protect\citeauthoryear{Li, Ren, Chen, Ren, Lian, and Ma}{Li
  et~al\mbox{.}}{2017}]%
        {li2017neural}
\bibfield{author}{\bibinfo{person}{Jing Li}, \bibinfo{person}{Pengjie Ren},
  \bibinfo{person}{Zhumin Chen}, \bibinfo{person}{Zhaochun Ren},
  \bibinfo{person}{Tao Lian}, {and} \bibinfo{person}{Jun Ma}.}
  \bibinfo{year}{2017}\natexlab{}.
\newblock \showarticletitle{Neural attentive session-based recommendation}. In
  \bibinfo{booktitle}{\emph{Proceedings of the 2017 ACM on Conference on
  Information and Knowledge Management}}. ACM, \bibinfo{pages}{1419--1428}.
\newblock


\bibitem[\protect\citeauthoryear{Liang, Charlin, and Blei}{Liang
  et~al\mbox{.}}{2016a}]%
        {liang2016causal}
\bibfield{author}{\bibinfo{person}{Dawen Liang}, \bibinfo{person}{Laurent
  Charlin}, {and} \bibinfo{person}{David~M Blei}.}
  \bibinfo{year}{2016}\natexlab{a}.
\newblock \showarticletitle{Causal Inference for Recommendation}. In
  \bibinfo{booktitle}{\emph{Causation: Foundation to Application, Workshop at
  UAI}}.
\newblock


\bibitem[\protect\citeauthoryear{Liang, Charlin, McInerney, and Blei}{Liang
  et~al\mbox{.}}{2016b}]%
        {liang2016modeling}
\bibfield{author}{\bibinfo{person}{Dawen Liang}, \bibinfo{person}{Laurent
  Charlin}, \bibinfo{person}{James McInerney}, {and} \bibinfo{person}{David~M
  Blei}.} \bibinfo{year}{2016}\natexlab{b}.
\newblock \showarticletitle{Modeling user exposure in recommendation}. In
  \bibinfo{booktitle}{\emph{Proceedings of the 25th International Conference on
  World Wide Web}}. \bibinfo{pages}{951--961}.
\newblock


\bibitem[\protect\citeauthoryear{Ma, Zhang, Cao, Jin, Wang, Liu, Ma, and
  Ren}{Ma et~al\mbox{.}}{2019}]%
        {ma2019jointly}
\bibfield{author}{\bibinfo{person}{Weizhi Ma}, \bibinfo{person}{Min Zhang},
  \bibinfo{person}{Yue Cao}, \bibinfo{person}{Woojeong Jin},
  \bibinfo{person}{Chenyang Wang}, \bibinfo{person}{Yiqun Liu},
  \bibinfo{person}{Shaoping Ma}, {and} \bibinfo{person}{Xiang Ren}.}
  \bibinfo{year}{2019}\natexlab{}.
\newblock \showarticletitle{Jointly learning explainable rules for
  recommendation with knowledge graph}. In \bibinfo{booktitle}{\emph{The World
  Wide Web Conference}}. \bibinfo{pages}{1210--1221}.
\newblock


\bibitem[\protect\citeauthoryear{McInerney, Lacker, Hansen, Higley, Bouchard,
  Gruson, and Mehrotra}{McInerney et~al\mbox{.}}{2018}]%
        {mcinerney2018explore}
\bibfield{author}{\bibinfo{person}{James McInerney}, \bibinfo{person}{Benjamin
  Lacker}, \bibinfo{person}{Samantha Hansen}, \bibinfo{person}{Karl Higley},
  \bibinfo{person}{Hugues Bouchard}, \bibinfo{person}{Alois Gruson}, {and}
  \bibinfo{person}{Rishabh Mehrotra}.} \bibinfo{year}{2018}\natexlab{}.
\newblock \showarticletitle{Explore, exploit, and explain: personalizing
  explainable recommendations with bandits}. In
  \bibinfo{booktitle}{\emph{Proceedings of the 12th ACM Conference on
  Recommender Systems}}. ACM, \bibinfo{pages}{31--39}.
\newblock


\bibitem[\protect\citeauthoryear{Peake and Wang}{Peake and Wang}{2018}]%
        {peake2018explanation}
\bibfield{author}{\bibinfo{person}{Georgina Peake} {and} \bibinfo{person}{Jun
  Wang}.} \bibinfo{year}{2018}\natexlab{}.
\newblock \showarticletitle{Explanation mining: Post hoc interpretability of
  latent factor models for recommendation systems}. In
  \bibinfo{booktitle}{\emph{Proceedings of the 24th ACM SIGKDD International
  Conference on Knowledge Discovery \& Data Mining}}.
\newblock


\bibitem[\protect\citeauthoryear{Pearl}{Pearl}{2000}]%
        {pearl2000causality}
\bibfield{author}{\bibinfo{person}{Judea Pearl}.}
  \bibinfo{year}{2000}\natexlab{}.
\newblock \bibinfo{booktitle}{\emph{Causality: models, reasoning and
  inference}}. Vol.~\bibinfo{volume}{29}.
\newblock \bibinfo{publisher}{Springer}.
\newblock


\bibitem[\protect\citeauthoryear{Rendle, Freudenthaler, Gantner, and
  Schmidt-Thieme}{Rendle et~al\mbox{.}}{2012}]%
        {rendle2012bpr}
\bibfield{author}{\bibinfo{person}{Steffen Rendle}, \bibinfo{person}{Christoph
  Freudenthaler}, \bibinfo{person}{Zeno Gantner}, {and} \bibinfo{person}{Lars
  Schmidt-Thieme}.} \bibinfo{year}{2012}\natexlab{}.
\newblock \showarticletitle{BPR: Bayesian personalized ranking from implicit
  feedback}.
\newblock \bibinfo{journal}{\emph{UAI}} (\bibinfo{year}{2012}).
\newblock


\bibitem[\protect\citeauthoryear{Rendle, Freudenthaler, and
  Schmidt-Thieme}{Rendle et~al\mbox{.}}{2010}]%
        {rendle2010factorizing}
\bibfield{author}{\bibinfo{person}{Steffen Rendle}, \bibinfo{person}{Christoph
  Freudenthaler}, {and} \bibinfo{person}{Lars Schmidt-Thieme}.}
  \bibinfo{year}{2010}\natexlab{}.
\newblock \showarticletitle{Factorizing personalized markov chains for
  next-basket recommendation}. In \bibinfo{booktitle}{\emph{Proceedings of the
  19th international conference on World wide web}}. ACM,
  \bibinfo{pages}{811--820}.
\newblock


\bibitem[\protect\citeauthoryear{Seo, Huang, Yang, and Liu}{Seo
  et~al\mbox{.}}{2017}]%
        {seo2017interpretable}
\bibfield{author}{\bibinfo{person}{Sungyong Seo}, \bibinfo{person}{Jing Huang},
  \bibinfo{person}{Hao Yang}, {and} \bibinfo{person}{Yan Liu}.}
  \bibinfo{year}{2017}\natexlab{}.
\newblock \showarticletitle{Interpretable convolutional neural networks with
  dual local and global attention for review rating prediction}. In
  \bibinfo{booktitle}{\emph{Proceedings of the Eleventh ACM Conference on
  Recommender Systems}}. \bibinfo{pages}{297--305}.
\newblock


\bibitem[\protect\citeauthoryear{Tang and Wang}{Tang and Wang}{2018}]%
        {tang2018personalized}
\bibfield{author}{\bibinfo{person}{Jiaxi Tang} {and} \bibinfo{person}{Ke
  Wang}.} \bibinfo{year}{2018}\natexlab{}.
\newblock \showarticletitle{Personalized top-n sequential recommendation via
  convolutional sequence embedding}. In \bibinfo{booktitle}{\emph{Proceedings
  of the Eleventh ACM International Conference on Web Search and Data Mining}}.
  ACM, \bibinfo{pages}{565--573}.
\newblock


\bibitem[\protect\citeauthoryear{Theodorou, Wortham, and Bryson}{Theodorou
  et~al\mbox{.}}{2017}]%
        {theodorou2017designing}
\bibfield{author}{\bibinfo{person}{Andreas Theodorou},
  \bibinfo{person}{Robert~H Wortham}, {and} \bibinfo{person}{Joanna~J Bryson}.}
  \bibinfo{year}{2017}\natexlab{}.
\newblock \showarticletitle{Designing and implementing transparency for real
  time inspection of autonomous robots}.
\newblock \bibinfo{journal}{\emph{Connection Science}} \bibinfo{volume}{29},
  \bibinfo{number}{3} (\bibinfo{year}{2017}), \bibinfo{pages}{230--241}.
\newblock


\bibitem[\protect\citeauthoryear{Wang, Wang, Jia, and Yin}{Wang
  et~al\mbox{.}}{2018b}]%
        {wang2018explainable}
\bibfield{author}{\bibinfo{person}{Nan Wang}, \bibinfo{person}{Hongning Wang},
  \bibinfo{person}{Yiling Jia}, {and} \bibinfo{person}{Yue Yin}.}
  \bibinfo{year}{2018}\natexlab{b}.
\newblock \showarticletitle{Explainable recommendation via multi-task learning
  in opinionated text data}. In \bibinfo{booktitle}{\emph{The 41st
  International ACM SIGIR Conference on Research \& Development in Information
  Retrieval}}. ACM, \bibinfo{pages}{165--174}.
\newblock


\bibitem[\protect\citeauthoryear{Wang, Guo, Lan, Xu, Wan, and Cheng}{Wang
  et~al\mbox{.}}{2015}]%
        {wang2015learning}
\bibfield{author}{\bibinfo{person}{Pengfei Wang}, \bibinfo{person}{Jiafeng
  Guo}, \bibinfo{person}{Yanyan Lan}, \bibinfo{person}{Jun Xu},
  \bibinfo{person}{Shengxian Wan}, {and} \bibinfo{person}{Xueqi Cheng}.}
  \bibinfo{year}{2015}\natexlab{}.
\newblock \showarticletitle{Learning hierarchical representation model for
  nextbasket recommendation}. In \bibinfo{booktitle}{\emph{Proceedings of the
  38th International ACM SIGIR conference on Research and Development in
  Information Retrieval}}. ACM, \bibinfo{pages}{403--412}.
\newblock


\bibitem[\protect\citeauthoryear{Wang, Chen, Yang, Wu, Wu, and Xie}{Wang
  et~al\mbox{.}}{2018a}]%
        {wang2018reinforcement}
\bibfield{author}{\bibinfo{person}{Xiting Wang}, \bibinfo{person}{Yiru Chen},
  \bibinfo{person}{Jie Yang}, \bibinfo{person}{Le Wu},
  \bibinfo{person}{Zhengtao Wu}, {and} \bibinfo{person}{Xing Xie}.}
  \bibinfo{year}{2018}\natexlab{a}.
\newblock \showarticletitle{A Reinforcement Learning Framework for Explainable
  Recommendation}. In \bibinfo{booktitle}{\emph{2018 IEEE International
  Conference on Data Mining (ICDM)}}. IEEE, \bibinfo{pages}{587--596}.
\newblock


\bibitem[\protect\citeauthoryear{Wang, Wang, Xu, He, Cao, and Chua}{Wang
  et~al\mbox{.}}{2019}]%
        {wang2019explainable}
\bibfield{author}{\bibinfo{person}{Xiang Wang}, \bibinfo{person}{Dingxian
  Wang}, \bibinfo{person}{Canran Xu}, \bibinfo{person}{Xiangnan He},
  \bibinfo{person}{Yixin Cao}, {and} \bibinfo{person}{Tat-Seng Chua}.}
  \bibinfo{year}{2019}\natexlab{}.
\newblock \showarticletitle{Explainable reasoning over knowledge graphs for
  recommendation}. In \bibinfo{booktitle}{\emph{Proceedings of the AAAI
  Conference on Artificial Intelligence}}, Vol.~\bibinfo{volume}{33}.
  \bibinfo{pages}{5329--5336}.
\newblock


\bibitem[\protect\citeauthoryear{Wood-Doughty, Shpitser, and
  Dredze}{Wood-Doughty et~al\mbox{.}}{2018}]%
        {wood2018challenges}
\bibfield{author}{\bibinfo{person}{Zach Wood-Doughty}, \bibinfo{person}{Ilya
  Shpitser}, {and} \bibinfo{person}{Mark Dredze}.}
  \bibinfo{year}{2018}\natexlab{}.
\newblock \showarticletitle{Challenges of Using Text Classifiers for Causal
  Inference}. In \bibinfo{booktitle}{\emph{Proceedings of the 2018 Conference
  on Empirical Methods in Natural Language Processing}}.
  \bibinfo{pages}{4586--4598}.
\newblock


\bibitem[\protect\citeauthoryear{Xian, Fu, Muthukrishnan, De~Melo, and
  Zhang}{Xian et~al\mbox{.}}{2019}]%
        {xian2019reinforcement}
\bibfield{author}{\bibinfo{person}{Yikun Xian}, \bibinfo{person}{Zuohui Fu},
  \bibinfo{person}{S Muthukrishnan}, \bibinfo{person}{Gerard De~Melo}, {and}
  \bibinfo{person}{Yongfeng Zhang}.} \bibinfo{year}{2019}\natexlab{}.
\newblock \showarticletitle{Reinforcement knowledge graph reasoning for
  explainable recommendation}. In \bibinfo{booktitle}{\emph{Proceedings of the
  42nd International ACM SIGIR Conference on Research and Development in
  Information Retrieval}}. \bibinfo{pages}{285--294}.
\newblock


\bibitem[\protect\citeauthoryear{Yu, Liu, Wu, Wang, and Tan}{Yu
  et~al\mbox{.}}{2016}]%
        {yu2016dynamic}
\bibfield{author}{\bibinfo{person}{Feng Yu}, \bibinfo{person}{Qiang Liu},
  \bibinfo{person}{Shu Wu}, \bibinfo{person}{Liang Wang}, {and}
  \bibinfo{person}{Tieniu Tan}.} \bibinfo{year}{2016}\natexlab{}.
\newblock \showarticletitle{A dynamic recurrent model for next basket
  recommendation}. In \bibinfo{booktitle}{\emph{Proceedings of the 39th
  International ACM SIGIR conference on Research and Development in Information
  Retrieval}}. ACM, \bibinfo{pages}{729--732}.
\newblock


\bibitem[\protect\citeauthoryear{Zhang and Chen}{Zhang and Chen}{2020}]%
        {zhang2018explainable}
\bibfield{author}{\bibinfo{person}{Yongfeng Zhang} {and} \bibinfo{person}{Xu
  Chen}.} \bibinfo{year}{2020}\natexlab{}.
\newblock \showarticletitle{Explainable Recommendation: A Survey and New
  Perspectives}.
\newblock \bibinfo{journal}{\emph{Foundations and Trends{\textregistered} in
  Information Retrieval}} (\bibinfo{year}{2020}).
\newblock


\bibitem[\protect\citeauthoryear{Zhang, Lai, Zhang, Zhang, Liu, and Ma}{Zhang
  et~al\mbox{.}}{2014}]%
        {zhang2014explicit}
\bibfield{author}{\bibinfo{person}{Yongfeng Zhang}, \bibinfo{person}{Guokun
  Lai}, \bibinfo{person}{Min Zhang}, \bibinfo{person}{Yi Zhang},
  \bibinfo{person}{Yiqun Liu}, {and} \bibinfo{person}{Shaoping Ma}.}
  \bibinfo{year}{2014}\natexlab{}.
\newblock \showarticletitle{Explicit factor models for explainable
  recommendation based on phrase-level sentiment analysis}. In
  \bibinfo{booktitle}{\emph{Proceedings of the 37th international ACM SIGIR
  conference on Research \& development in information retrieval}}. ACM,
  \bibinfo{pages}{83--92}.
\newblock


\end{thebibliography}

\appendix

\end{document}